# To Explore the Potential Inhibitors against Multi-target Proteins of COVID-19 using In-Silico Study

Imra Aqeel [1]

1   Biomedical Informatics Research Lab, Department of Computer & Information Sciences, Pakistan Institute of Engineering & Applied Sciences, Nilore, Islamabad 45650, Pakistan; imraaqeel@pieas.edu.pk ; abdulmajiid@pieas.edu.pk

**Abstract:**

The global pandemic due to emergence of COVID-19 has created the unrivaled public–c health crisis. It has huge morbidity rate never comprehended in the recent decades. Researchers have made many efforts to find the optimal solution of this pandemic. Progressively, drug repurposing is an emergent and powerful strategy with saving cost, time, and labor. Lacking of identified repurposed drug candidates against COVID-19 demands more efforts to explore the potential inhibitors for effective cure. In this study, we used the combination of molecular docking and machine learning regression approaches to explore the potential inhibitors for the treatment of COVID-19. The 5903 approved drugs from ZINC database, have screened against the multi-target proteins of COVID-19 including *Spike (S), main protease 3CL (3CLpro),* and *Nucleocapsid (N)* proteins. These proteins are responsible for binding of virons with host cell receptors, replicating the virus, and wrapping the virus RNA into a helical symmetrical structure. We calculated the binding affinities of these drugs to multi-target proteins using molecular docking process. We perform the QSAR modeling by employing various machine learning regression approaches to identify the potential inhibitors against COVID-19. Our findings with best scores of $R^2$ and RMSE demonstrated that our proposed Decision Tree Regression (DTR) model is the most appropriate model to explore the potential inhibitors. We proposed five novel promising inhibitors with their respective Zinc IDs *ZINC (3873365, 85432544, 8214470, 85536956, and 261494640)* within the range of -19.7 kcal/mol to -12.6 kcal/mol. We further analyzed the physiochemical and pharmacokinetic properties of these most potent inhibitors to examine their behavior. The analysis of these properties is the key factor to promote an effective cure for public health. Our work constructs an efficient structure with which to probe the potential inhibitors against COVID-19, creating the combination of molecular docking with machine learning regression approaches. Our findings contribute to the larger goal of finding effective cures for COVID-19, which is the acute global health challenge. Our research provides a foundation for improving the future response to viral threats and assists in managing the current pandemic and projecting future public health measures.

## 1- Introduction

According to https://www.worldometers.info/coronavirus/, as of July 10, 2024, there were over 704.75 million confirmed cases of COVID-19 worldwide, and over 7.01 million deaths from the virus. This has created an unparalleled global health emergency. The respiratory tract infections caused by the causative agent, Severe Acute Respiratory Syndrome Coronavirus 2 (SARS-CoV-2), range in severity from subclinical and mild common cold-like symptoms to fatal acute respiratory distress syndrome and multiple organ failure. SARS-CoV-2 is an enveloped single positive-stranded RNA virus that belongs to the Coronaviridae family [1]. SARS-CoV-2 attacks airway epithelium cells and is spread via aerosol. The severity of the disease has decreased as a result of vaccination campaigns and rising public immunity [2]. Nonetheless, the range of drugs available for COVID-19 treatment and prevention is restricted. Furthermore, there is still much to learn about the biology of SARS-CoV-2 infection.

Even while vaccinations have helped contain the Covid-19 outbreak, their full efficacy depends on herd immunity, a communal endeavor at odds with the population's well-known vaccine hesitation [3]. Furthermore, finding efficient antivirals is a top priority due to the virus's quick mutation, diminishing immunity, and expensive vaccination. Currently, there are few antivirals that are approved under emergency authorization but have low effectiveness against SARS-CoV-2 replication. The first approved antiviral *remdesivir*, [4] has a broad-spectrum antiviral activity with RNA polymerase inhibitor. It can only be used to treat a restricted number of hospitalized patients and requires parenteral administration. Oral nucleotide analog *molnupiravir* [5] has a broad-spectrum antiviral action and is used clinically in individuals with mild to moderate Covid-19 who are at high risk of developing severe illness. Lastly, the most popular *paxlovid* consists of a combination of *ritonavir*, a *CYP3A4* inhibitor, and *nimatrelvir*, the primary protease inhibitor for SARS-CoV-2. In addition to suppressing host or viral proteases or RNA synthesis, other methods for disrupting the coronavirus life cycle include stopping the virus's entry into cells through the ACE2 receptor, preventing the assembly of new viral particles, and obstructing the virus's uptake pathway. Modulating inflammatory pathways can also have an impact on the infection's outcome [6] [7]. Other already promoted antivirals like *lopinavir* or *ritonavir* finally shown fruitless to treat the disease [8].

The reasons listed above highlight the necessity of creating targeted antiviral treatments in the near future to combat SARS-CoV-2. However, the protracted process of developing new medications makes this urgency lessened [9]. Because drugs have well-established pharmacological qualities, repurposing them can assist reduce the time it takes to introduce a medicine into the clinic [10]. Conversely, the use of computational approaches can aid in the direction of discovery activities, saving money by eliminating the need for expensive trial and error procedures involving living systems, biochemical screening, and cell cultures [11]. For instance, *baricitinib*, a medication for rheumatoid arthritis, was repurposed using artificial intelligence approaches to treat SARS-CoV-2 infections [12]. This medication has been demonstrated to shorten recovery times and hasten the improvement of clinical status in Covid-19 patients when taken in conjunction with *remdesivir* [13].

Angiotensin-converting enzyme-2 (ACE2) receptor has to attach to the host cell, and this is seen by an extremely glycosylated envelope protein known as *spike or S* protein that is used by both SARS-COV-2 and SARS-COV [14] [15]. The S1 and S2 subunits combined to form the S protein. The earlier of which is used to bind to the ACE2 receptor and the latter of which intricate in cell infection and membrane fusion, triggered by the serine protease TMPRSS2 [16] [17]. Additionally, the receptor binding domain (RBD), is encoded by the S1 subunit, is essential to attach to the ACE2 peptidase domain and it is the primary target of deactivating the antibodies produced during infection. [18][19].

The other viral protein of SARS-COV-2 is *nucleocapsid (N)* protein which is vital to the coronavirus life cycle and involved in numerous crucial activities after virus incursion. It identifies and surrounds the viral RNA forming a helical symmetrical structure that plays essential multiple function in the life cycle of coronavirus [20]. It attaches itself to the virus's genomic RNA to create a ribonucleoprotein complex (RNP) [21]. It has garnered a lot of interest in the creation of drugs and vaccines.

SARS-CoV-2 requires a variety of viral proteins to aid in its reproduction, but one of these proteins - the main protease *3CLpro*, also known as main protease *(Mpro)*, is essential for cleaving viral polyproteins into the functional non-structural proteins required for viral replication [22]. Due to its importance in the viral life cycle, *3CLpro* is now being considered as a possible target for COVID-19 antiviral treatment development. The homodimeric cysteine protease 3CL protease is a desirable option for the creation of protease inhibitors due to the catalytic dyad of *His41* and *Cys145* [23]. Numerous investigations have documented the accomplishment of pinpointing peptides and tiny compounds that can potently impede the *3CL protease*'s activity in vitro [24]. The creation of precise and effective *3CL protease* inhibitors is still a difficult task.

However, the complexity and fast evolution of the disease pose huge challenges in identifying promising repurposed drugs as a potential treatment for COVID-19. For example, molecular docking [25] is one of the computational approaches that has proved to be informative on potential drug candidates in the identification of possible drug candidates because it allows quick screening of large compound numbers against specific targets, producing valuable information regarding their potential efficacy and binding interactions. These types of studies open possibilities on the identification of computationally obtained compounds as new drugs for the treatment of COVID-19, but they also highlight the need for a machine learning-based framework to speed up the process.

In this paper, we outline our strategy for investigating Covid-19 infection inhibitors through the establishment of a molecular docking and drug screening machine learning study. We present the findings of a virtual screening investigation aimed at detecting hits that bind to SARS-CoV-2 multi-proteins, S, N, and Mpro, which inhibit entry and viral engagement. A virtual screening was conducted with the Zinc database of authorized medications. The machine learning QSAR modeling and molecular docking were considered in the computational process. The autodock vina software [26] was utilized to conduct molecular docking studies based on the various structures that were found. Using ever stricter computational criteria, the molecular docking data were sifted via a virtual screening process that resulted in the reduction of the original set of 5903 compounds to a final list of 5537 molecules. After evaluating the top five hits in silico for their possible ability to impede the Covid-19 interaction, a shortlist of 05 compounds is created. Our goal was to identify inhibitors that are selective and concentrate on particular interactions that enhance the binding of the SARS-CoV-2 multi-targeted proteins.

In our proposed study, to create a more effective method of estimating binding affinities towards the multi-targeted proteins of Covid-19, we employed QSAR modeling with multiple ML techniques. Several regression models, including Decision Tree regression (DTR), Gradient Boosting regression (GBR), Extra Trees regression (ETR), K-Nearest Neighbor regression (KNNR), Multi-Layer Perceptron regression (MLPR), and XGBoost regression (XGBR), are constructed and trained using MACCS fingerprint feature descriptor type—that is computed using the PaDEL descriptor program [27]. To enhance the performance of regression models, the input dataset is split into two sections, with 80% of the internal dataset being used for 5-fold cross validation. To test these models, an external dataset including the remaining 20% of the data is employed. $R^2$ and RMSE are statistical measurements that are used to evaluate the simulated outcomes of regression models. Comparing our intended DTR model to other regression models, we found that its $R^2$ and RMSE values were better. This demonstrates that the DTR model works well for predicting binding affinities. Finally, we conducted a physiochemical study of shortlisted inhibitors with the help of in silico analyses in order to comprehend how medication molecules behave in biological systems and assess the safety and efficacy of these compounds.

This paper's subsequent sections are organized as follows: The results and the discussion that went along with them are shown in section 2. The material and proposed computational framework used in this investigation are described in section 3. Section 4 presents the findings and conclusions from our study.

## 2- Results and Discussion

In order to investigate possible inhibitors against COVID-19 through drug repurposing strategies and forecast their binding affinities for approved drugs towards the multi-target proteins *Spike, 3CLpro,* and *nucleocapsid,* we constructed a number of machine learning (ML) based QSAR models in this work. Using MACCS fingerprint feature set, we will first evaluate how well our suggested model DTR predicts binding affinities. Next, using the crucial statistical metrics of $R^2$ and RMSE, compare its performance with a number of QSAR models. The outcomes of the approved drugs' molecular docking studies and their interactions with multi-target proteins will next be discussed. In order to determine the drugs' safety and effectiveness towards the multi-target proteins, we will lastly perform a physiochemical investigation of the drug compounds that made the short list.

### *2.1. Evaluation of QSAR Model*

The present study suggests a process for building a QSAR model that utilizes the Decision Tree Regression (DTR) technique. A detailed explanation of the 4639 drug compounds used in the development of the model may be found in section 3.2.1. As mentioned in section 3.2.2, two different feature sets are utilized to assess the performance of the proposed model: fingerprint features are inserted into the X matrix to create the data matrices, and their matching binding affinities are placed in the Y matrix. The dataset is split into two parts: an internal dataset, which makes up 80% of the total, and an external dataset, which makes up 20%. We trained and validated the model based on an internal dataset using 5-fold cross validation, and now we assess the performance of the model on the external dataset.

We utilize $R^2$ and RMSE, two widely recognized statistical measures, to evaluate the effectiveness of our suggested QSAR model. $R^2$ measures the degree to which the independent variables (features) account for the variation in the dependent variable (binding affinity) and assesses the model's fitness. Higher values indicate greater model performance; the range is 0 to 1. On the other hand, the relative error between the actual and predicted values of binding affinity is measured by RMSE. For MACCS fingerprint feature set towards multi-target proteins *Spike, 3CLpro,* and *nucleocapsid*, **Figure 1** shows the actual and predicted binding affinities to display the efficacy of the DTR model.

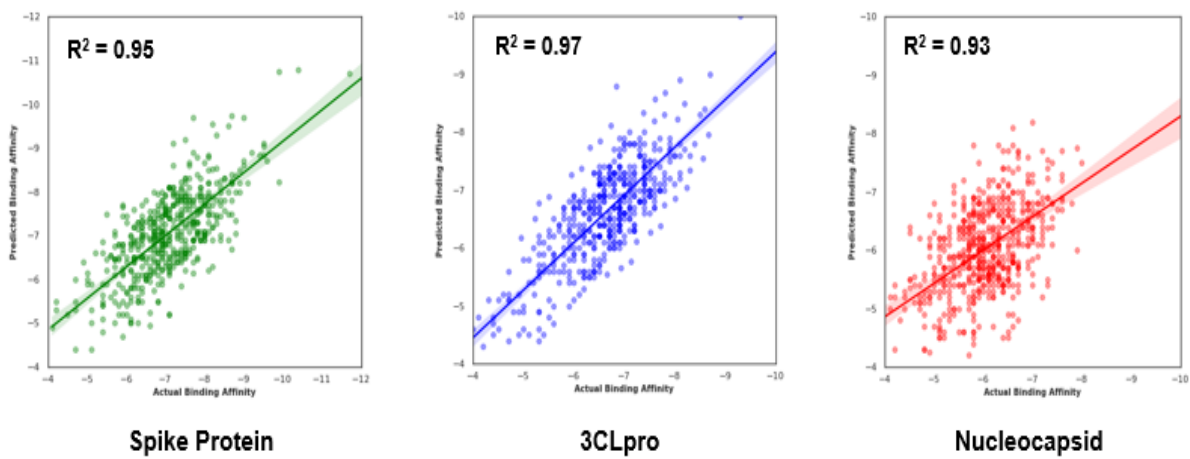

**Figure 1. Regression plots of $R^2$ using MACCS fingerprint features for external dataset.**

In this study, MACCS fingerprint feature set is used to assess the DTR model's efficacy. The evaluation results are presented in **Table 1** along with the $R^2$, Mean Squared Error (MSE), and RMSE values.

**Table 1. Performance of DTR Model for MACCS Fingerprint.**

| Sr. No | Protein Name | $R^2$ | MSE | RMSE |
|---|---|---|---|---|
| 1 | Spike | 0.95 | 2.76 | 1.66 |
| 2 | 3CLpro | 0.97 | 2.46 | 1.57 |
| 3 | Nucleocapsid | 0.93 | 2.21 | 1.49 |

**Table 1** shows the effectiveness of the Decision Tree Regression (DTR) model in predicting binding affinities across three important SARS-CoV-2 proteins: *Spike, 3CLpro,* and *Nucleocapsid*, using the MACCS fingerprint. The following performance metrics, such as the coefficient of determination ($R^2$), the mean square error (MSE), and the root mean square error (RMSE) are employed. How much variance in the predictions of binding affinities can explained by the model is represented by $R^2$ measure. The model attained strong $R^2$ values for all three proteins: 0.95 for *Spike*, 0.97 for *3CLpro*, and 0.95 for *Nucleocapsid*. These outcomes specified the high performance for each target and suggest that our proposed DTR model elucidates most of the variance in the predictions.

Other details in relation to the accuracy of the model further described by MSE and RMSE. Thus, according to both MSE and RMSE indices, the *3CLpro* shares the best accuracy among the three proteins, with MSE of 2.46 and RMSE of 1.57. This is also evidenced by the model of the *Spike* protein, which achieved MSE of 2.76 as well as RMSE of 1.66. In comparison to *Spike*, the *Nucleocapsid* protein model shows somewhat higher prediction performance, with the lowest MSE of 2.21 and RMSE of 1.49. The DTR model is robust in predicting binding affinities using MACCS fingerprints; overall, it works well for all proteins, with minimal error margins in predictions.

In QSAR modeling, comparing $R^2$ and $Q^2$ values is the most used metric for evaluating performance. Less than 0.3 should separate them [28]. Additionally, the model performs well in regression with a $Q^2$ score of 0.5, while outstanding performance is indicated by a value above 0.9. For *spike* protein, our model produced an $R^2$ value of 0.80 for training data and a $Q^2$ value of 0.95 for testing data. For *3CLpro* and *nucleocapsid*, the corresponding values are 0.77, 0.97; and 0.77, 0.93 for $R^2$ and $Q^2$, respectively. There's a tiny difference between $R^2$ and $Q^2$ in the range of 0.15 - 0.20. This suggests that, with adequate binding affinity estimate power, the suggested QSAR prediction model is the most appropriate.

## 2.2. Comparative Analysis

We use the MACCS fingerprint feature descriptor to compare our proposed QSAR model with different regression models in order to evaluate its accuracy and predictive performance. $R^2$ and RMSE are statistical measures that are employed as assessors.

As we proposed our DTR model, the performance of
**Table 2** is a comparison with regression models based on the MACCS fingerprint feature descriptor for the DTR model, regarding the values of $R^2$ and RMSE on the MACCS Fingerprint.

**Table 2. Performance comparison of QSAR regression models on MACCS fingerprint.**

| Regression Model | Spike | | 3CLpro | | Nucleocapsid | |
|---|---|---|---|---|---|---|
| | $R^2$ | RMSE | $R^2$ | RMSE | $R^2$ | RMSE |
| DTR | 0.95 | 1.66 | 0.97 | 1.57 | 0.93 | 1.49 |
| ETR | 0.79 | 1.80 | 0.85 | 1.80 | 0.70 | 1.63 |
| KNNR | 0.66 | 1.82 | 0.96 | 1.86 | 0.87 | 1.71 |
| GBR | 0.70 | 1.84 | 0.88 | 1.81 | 0.69 | 1.62 |
| MLPR | 0.61 | 1.74 | 0.60 | 1.64 | 0.65 | 1.61 |
| XGBR | 0.52 | 1.85 | 0.54 | 1.85 | 0.52 | 1.73 |

**Table 2** presents a comparison of several QSAR regression models in predicting the affinities of binding for three major proteins associated with SARS-CoV-2: *Spike, 3CLpro,* and *Nucleocapsid*, employing MACCS fingerprint for Decision Tree Regression (DTR), Extra Trees Regression (ETR), K-Nearest Neighbors Regression (KNNR), Gradient Boosting Regression (GBR), Multi-Layer Perceptron Regression (MLPR), and XGBoost Regression (XGBR). **R²** and RMSE are the evaluation measures used for the models.

Our proposed DTR model consistently exhibits the best predictive performance across all three proteins, with **R²** values of 0.95, 0.97, and 0.93 for *Spike, 3CLpro,* and *Nucleocapsid*, respectively, along with the lowest RMSE values of 1.66, 1.57, and 1.49. This indicates that the DTR model explains the largest proportion of variance in the binding affinity predictions and has the most accurate predictions among the models tested.

The ETR model also performs relatively well, particularly for *3CLpro*, with an R² of 0.85 and an RMSE of 1.80. However, its performance declines for *Spike* (R² = 0.79) and *Nucleocapsid* (R² = 0.70), with RMSE values slightly higher at 1.80 and 1.63, respectively.

KNNR achieves good results for *3CLpro* with an R² of 0.96 and an RMSE of 1.86 but shows lower predictive performance for *Spike* (R² = 0.66) and *Nucleocapsid* (R² = 0.87), with RMSE values of 1.82 and 1.71. GBR displays moderate performance, with R² values of 0.70 for *Spike*, 0.88 for *3CLpro*, and 0.69 for *Nucleocapsid*, and RMSE values slightly higher than those of KNNR.

The MLPR and XGBR models perform poorly across all proteins, with R² values ranging from 0.52 to 0.65 and RMSE values consistently higher than those of the top-performing models. XGBR appears to have the weakest predictive ability for all three proteins with R² values of 0.52 for both *Spike* and *Nucleocapsid* and 0.54 for *3CLpro*.

Overall, the results indicate that DTR outperforms the other regression models in predicting binding affinities across all three proteins, with the lowest RMSE and highest R² values. This suggests that DTR is the most suitable model for this specific task when using MACCS fingerprints, while models like MLPR and XGBR show significantly weaker performance.

## 2.3. Molecular Docking

Determining how well the chosen therapeutic compounds interacted with three important target proteins was our main goal. In order to anticipate and examine the interactions between ligands and target proteins, the molecular docking technique is employed. It facilitates comprehension of the orientation and binding affinity of ligands within the active site of proteins. In this work, we estimated the binding affinities of drug compounds taken from the Zinc database using a ligand-based docking technique. After the drug compounds were formatted in PDBQT, the binding affinities between them and the target proteins were calculated and expressed in kcal/mol.

The three-dimensional interaction image of the target protein *7LM9* complex, which has ligand *2297* attached to it, is shown in **Figure 2 (a). Figure 2 (b),** on the other hand, shows the three-dimensional interaction image of the target protein *7JSU* complex when ligand *2297* is coupled to it. However, the 3D interaction view of the target protein *7DE1* complex with ligand *2297* attached to it is shown in **Figure 2 (c).**

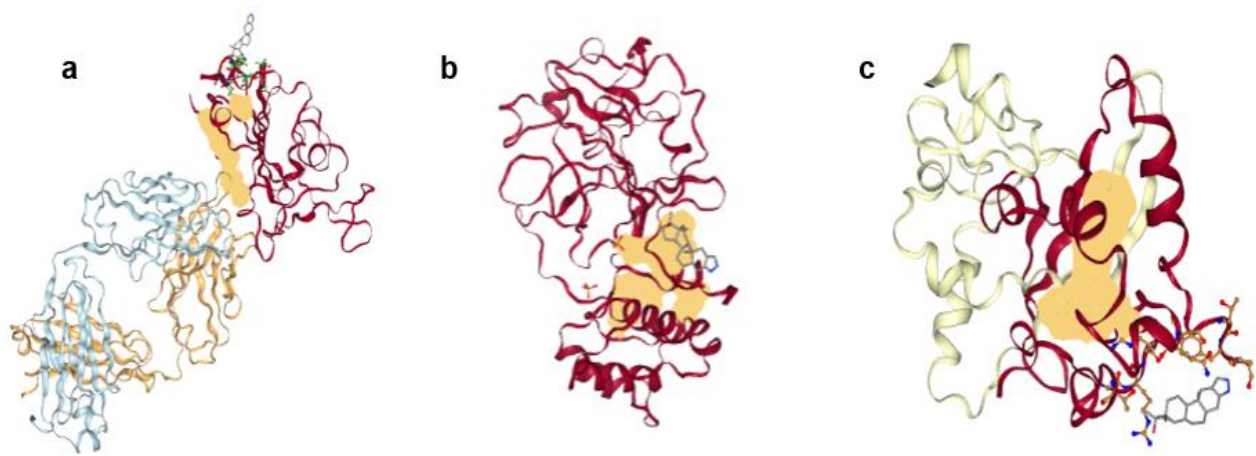

**Figure 2. 3D interaction view of proteins, (a) 7LM9, (b) 7JSU, (c) 7DE1, with ligand ID 2297.**

Another two measurements used to evaluate proteins for drug purposes and biotechnological applications are the Drug Score and the Simple Score. The Drug Score assesses ability of a protein to be targeted by small molecules based protein structural features that favor drug targeting, drug similarity, and binding affinity. The greater Drug Score a protein carries the higher chances it has of being a therapeutic target. However, the Simple Score presents the general properties of the protein more directly, taking into account such characteristics as hydrophobicity, molecular weight, and content of particular amino acids. For instance, constancy in industrial processes or enzyme activity, a high Simple Score demonstrates proper features for operation and performance in different applications. Altogether, these scores give valuable information for the scientists who work in the field of the drug creation and proteins modification.

**Table *3*** presents data on three target proteins (*7LM9, 7JSU,* and *7DE1*) and their ligand ID *2297* and the binding pockets associated with this ligand. Significant data that are useful for assessing the suitability of these proteins for drug targeting are presented.

**Table 3. Drug Score of Target proteins with Ligand ID 2297.**

| Protein Name | Pocket Name | Volume Å³ | Surface Å² | Drug Score | Simple Score |
|---|---|---|---|---|---|
| 7LM9 | P_0 | 535.49 | 591.53 | 0.87 | 0.35 |
| 7JSU | P_0 | 746.24 | 1257.1 | 0.82 | 0.54 |
| 7DE1 | P_0 | 525.18 | 161.17 | 0.74 | 0.38 |

To facilitate the analysis of binding properties, **Table 3** with most suitable binding pocket, P_0, for each protein. The binding pocket volume is expressed in cubic angstroms (Å³), which quantifies the potential capacity for bound ligands. For example, a 535.49 Å³ *7LM9* binds ligand ID *2297* with high affinity. Another important measure of understanding interaction sites is the solvent-accessible surface area, again measured in square angstroms (Å²). *7LM9* has a surface area of 591.53 Å² suggesting a good region for ligand binding. The score is 0.87, that is, very potent binding, thus, *7LM9* stands exceptional from the rest with a Drug Score, which is the possibility to bind effectively with ligand ID *2297*. Comparatively, *7DE1* has a score of 0.74 and *7JSU* has its Drug Score at 0.82 showing greater possibilities for favorable interactions. Moreover, for the Simple Score, we checked for the basic functional properties of the binding pockets. For *7LM9, 7JSU,* and *7DE1*, the scores were 0.35, 0.54, and 0.38, respectively, giving us a little better idea about what makes them unique. All of these values emphasize the benefits of each protein in favor of ligand ID *2297*, making their use highly relevant to drug discovery projects.

On the other hand, the root mean square deviation (RMSD) of the ligand from the starting site of the protein complex is used to evaluate the accuracy of docking. A ligand molecule with a better docking geometry has a lower RMSD value. Remarkably, our study shows that in their ideal positions, each of the five ligands had an RMSD value of zero, suggesting that the docking geometry was highly accurate. The binding affinities (BA) values between the five drug compounds that score highest and the associated proteins are displayed in **Table *4***. The best BA values for the three important proteins, *7LM9, 7JSU,* and *7DE1*, were produced by the ligands *2297, 4434, 4440, 3172,* and *5471*. According to this table, the most promising drug molecule is ligand *2297*, which has minimum values of -19.7, -15.1, and -19.2 towards *7LM9, 7JSU,* and *7DE1*, respectively. The study of ligand poses and the associated binding energies contributes to the understanding of the connection between the ligands' molecular structure and binding affinity. It indicates that the ligand molecule has a strong connection and may be a promising treatment target.

**Table 4. Top ranked five ligands with target proteins.**

| Sr. No. | Zinc ID | Ligand ID | 7LM9 (BA) (kcal/mol) | 7JSU (BA) (kcal/mol) | 7DE1 (BA) (kcal/mol) |
|---|---|---|---|---|---|
| 1 | ZINC003873365 | 2297 | -19.7 | -15.1 | -19.2 |
| 2 | ZINC085432544 | 4434 | -15.3 | -14.4 | -14.0 |
| 3 | ZINC085536956 | 4440 | -14.4 | -13.9 | -12.6 |
| 4 | ZINC008214470 | 3172 | -15.3 | -13.6 | -14.3 |
| 5 | ZINC261494640 | 5471 | -13.9 | -13.6 | -14.4 |

## 2.4. Physiochemical analysis of drug candidates

We identified five therapeutic compounds with higher potency and robust interactions with three target proteins, *Spike, 3CLpro,* and *Nucleocapsid,* based on our investigation. A thorough description of these five therapeutic compounds with the lowest binding energies is described in

**Table 5.** The drug compounds' 2D structure, SMILES, molecular formula, and Zinc ID are all included in the description.

**Table 5. Description of top ranked five drug compounds.**

| Zinc ID | Molecular Formula | SMILES | 2D Structure |
|---|---|---|---|
| ZINC085432544 | C46H58N4O9 | CC[C@]1(O)C[C@@H]2CN(CCc3c([nH]c4ccccc34)[C@@](C(=O)OC)(c3cc4c(cc3OC)N(C)[C@H]3[C@@](O)(C(=O)OC)[C@H](OC(C)=O)[C@]5(CC)C=CCN6CC[C@]43[C@@H]65)C2)C1 | 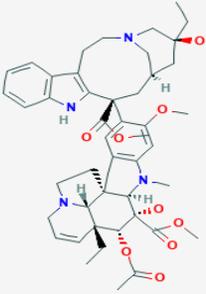 |
| ZINC003873365 | C21H32N2O | C[C@]12Cc3cn[nH]c3C[C@@H]1CC[C@H]1[C@@H]2CC[C@]2(C)[C@H]1CC[C@@]2(C)O | 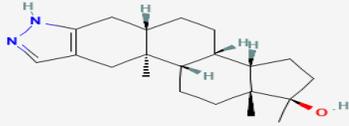 |
| ZINC085536956 | C45H54N4O8 | CCC1=C[C@H]2CN(C1)Cc1c([nH]c3ccccc13)[C@@](C(=O)OC)(c1cc3c(cc1OC)N(C)[C@H]1[C@@](O)(C(=O)OC)[C@H](OC(C)=O)[C@]4(CC)C=CCN5CC[C@]31[C@@H]54)C2 | 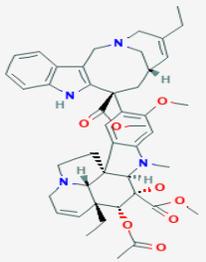 |
| ZINC261494640 | C50H77NO13 | CO[C@H]1C[C@@H]2CC[C@@H](C)[C@@](O)(O2)C(=O)C(=O)N2CCCC[C@H]2C(=O)O[C@H]([C@H](C)C[C@H]2CC[C@H](O)[C@@H](O)C2)CC(=O)[C@@H](C)/C=C(/C)[C@@H](O)[C@H](OC)C(=O)[C@H](C)C[C@H](C)/C=C\C=C/C=C\1C | 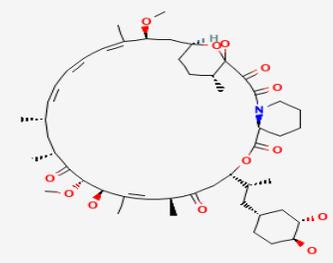 |
| ZINC00082 14470 | C43H55N5O7 | CC[C@]1(O)C[C@H]2CN(CCc3c([nH]c4ccccc34)[C@@](C(=O)OC)(c3cc4c(cc3OC)N(C)[C@H]3[C@@](O)(C(N)=O)[C@H](O)[C@]5(CC)C=CCN6CC[C@]43[C@@H]65)C2)C1 | 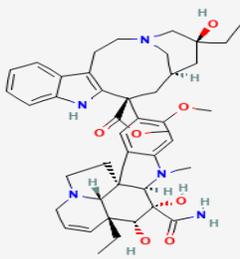 |

Table 6 provides a group of some physicochemical properties of five potential therapeutic possibilities identified by their Zinc ID. The most influencing physicochemical properties control the efficiency and bioactivity of medicinal drugs. Drug development captures the important attributes that accompany pharmacokinetic profiling and effectiveness. It is significant to establish the physicochemical properties of pharmaceutical compounds to gain knowledge in respect of their safety and effectiveness in biological systems. These include, among others, its molecular weight (Mol.Wt), LogP, rings, hydrogen bond donors (HBD), hydrogen bond acceptors (HBA), heavy atoms, heteroatoms, and sp3 fraction, all of which influence the binding affinity of therapeutic candidates towards specific target proteins.

For instance, molecular weight is regarded as one of the most critical criteria for the assessment of pharmacokinetics and pharmacodynamics properties of a drug or its penetration through cell membranes and interaction with target proteins. Optimum molecular weight has been demonstrated to be crucial for increasing the chances of binding with the target proteins. LogP is another important constituent which has undergone rigorous roles in impacting the therapeutic binding affinity. Higher logP value implies higher lipophilicity and a high tendency to interact with the hydrophobic domains of proteins. LogP has been identified to be the key descriptor responsible for how well small molecule therapeutics can bind to their target in one study [29]. Moreover, HBD and HBA are also the background of drug-protein interactions in virtue of their contribution to the generation of specific molecular interactions relevant to binding and recognition. The number of rings in a pharmacological molecule is another very important characteristic, with influence on selectivity and binding affinity for particular targets. Rings that are too large or have too many rings may interfere with binding or increase toxicity, whereas drugs that contain two to three rings are more likely to show good oral bioavailability and target affinity [30]. Heavy and heteroatoms, particularly nitrogen and oxygen, specific drug-target interactions by forming hydrogen bonds or other electrostatic interactions with the target residues. However, an imbalance of too few or too many heteroatoms could affect membrane permeability, a medicine's solubility, and other essential components of bioavailability.

The vital characteristic, a drug compound's physicochemical characteristics and target affinity have been demonstrated to be influenced by its sp3 fraction [31]. An indicator of the proportion of carbon atoms with three or more single bonds is the sp3 fraction. Increased water solubility, decreased toxicity, and enhanced pharmacokinetic features are associated with greater sp3 fractions, whereas low levels may result in poor bioavailability or decreased target selectivity.

Table 6. Physiochemical properties of selected drug compounds.

| Sr. No | Zinc ID | Mol. Wt (g/mol) | Rings | LogP | HBD, HBA | Hetero Atoms | Heavy Atoms | Fraction sp3 |
|---|---|---|---|---|---|---|---|---|
| 1 | ZINC085432544 | 810.9 | 9 | 3.99 | 5, 10 | 13 | 59 | 0.59 |
| 2 | ZINC003873365 | 328.5 | 5 | 4.118 | 2, 2 | 3 | 24 | 0.86 |
| 3 | ZINC085536956 | 778.9 | 9 | 4.754 | 4, 9 | 12 | 57 | 0.53 |
| 4 | ZINC261494640 | 900.1 | 4 | 5.527 | 4, 10 | 14 | 64 | 0.74 |
| 5 | ZINC008214470 | 753.9 | 9 | 2.732 | 7, 8 | 12 | 55 | 0.58 |

The drugs' molecular weights, which vary widely in size from 328.5 g/mol to 900.1 g/mol in the table, show this. We noticed that there were generally more heavy and heteroatoms in the drugs with larger molecular weights. This suggests that bigger molecules might bind to particular targets more successfully. But we also noticed that a higher fraction sp3 was found in several of the drugs with lower molecular weights, which would indicate a higher level of three-dimensional complexity and possibly better binding interactions. Additionally, the logP values vary from 2.732 to 5.527, with the majority of drugs falling between 3 and 5. Drug development relies heavily on lipophilicity since it affects drug absorption, distribution, metabolism, and excretion (ADME) characteristics. In addition, the majority of the HBD and HBA readings in this table are $\geq 4$ and $\geq 8$, respectively. A drug molecule's capacity to create many hydrogen bonds with the protein binding site increases with increasing HBD and HBA levels. This may enhance the interaction and lead to a higher binding affinity.

The table indicates that the molecules have between 4 to 9 rings, and from 24 to 64 heavy atoms. The number of rings and heavy atoms within a molecule may influence its stability, potency, and selectivity. Typically, compounds containing several heavy atoms and multiple rings are much more complex and thus more likely to interact with the target receptor. However from this table, we have a heteroatom count that varies from 3 to 14 with most of them between 12 and 14, and these are also important, as they may give clues to the potential binding interactions that might occur between drug molecules and their specific target. The fraction sp3 is another significant characteristic that has a range between 0.53 and 0.86, representing percentage carbon atoms in the drug that are sp3 hybridized. Most drugs have a sp3 fraction value above 0.50. A drug whose sp3 value is significantly higher will exhibit more three-dimensional character and is consequently much more likely to be effective when binding to a particular target.

Generally, these physicochemical characteristics of the five chosen drug compounds are basic to the understanding and aiding in drug development. These are the most important attributes for drugs with enhanced pharmacokinetic characteristics and maximum efficacious properties. These drug compounds may be used again to treat different illnesses. Further in vitro and in vivo investigations are required to ascertain the safety, dose, and effectiveness of these compounds for the treatment of particular disorders. These physicochemical features might be helpful in these investigations. Research on drug repurposing can start with this insightful examination of these drugs. This demonstrates how important it is to take into account the physicochemical characteristics of the drugs while repurposing them. **Table 7** lists the generic name and original prescription for each of our recommended drug choices that we suggest to repurpose against COVID-19.

**Table 7. Suggested drugs for repurposing in this study.**

| Sr. No. | Zinc ID | Generic Name | Orignal Prescription | New Indication |
|---|---|---|---|---|
| 1 | ZINC003873365 | Stanozolol | Certain forms of breast cancer, anemia, and hereditary angioedema (HAE). | COVID-19 3CL |
| 2 | ZINC085432544 | Vinblastine | Breast cancer, neuroblastoma, Hodgkin's and non-Hodgkins lymphoma, testicular cancer, histiocytosis, mycosis fungoides and Kaposi's sarcoma. | COVID-19 3CL |
| 3 | ZINC085536956 | Vinorelbine tartrate | Advanced or metastatic non-small cell lung cancer. | COVID-19 3CL |
| 4 | ZINC008214470 | Vindesine | Malignant lymphoma, Acute leukaemia, Hodgkin's disease, acute erythraemia and acute panmyelosis | COVID-19 3CL |
| 5 | ZINC261494640 | 41-O-demethyl rapamycin | Prevent organ rejection in kidney transplant patients, tumor-based cancers, and coat stents implanted in heart disease patients | COVID-19 3CL |

## 3- Material and Methods

Three components make up the computational framework that this study proposes, as shown in **Figure 3**. Module A includes the methods necessary to extract approved drugs and target proteins from the corresponding databases and compute the binding affinities between the target proteins and the extracted drug molecules. However, module B details the development of the QSAR model and compares its efficacy with several cutting-edge machine learning algorithms. Finally, module C places a strong emphasis on drug analysis and molecular docking studies.

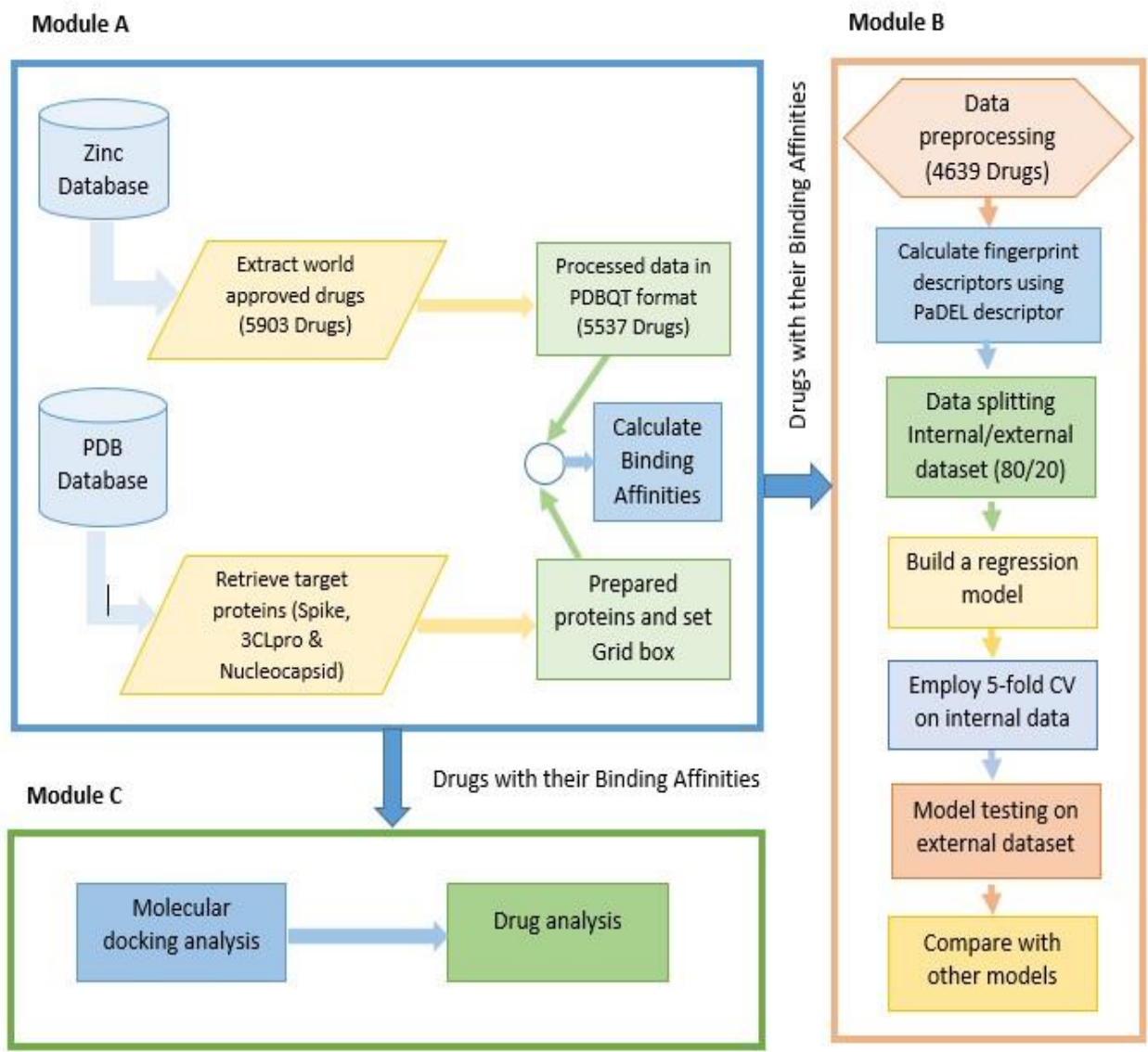

**Figure 3. Three principle modules (A to C) in the proposed computational framework.**

## 3.1. Module A: Dataset Preparation and Molecular Docking

Module A the stages of data preprocessing which are as follows :

### 3.1.1. Targeting the *Spike, 3CLpro,* and *Nucleocapsid* Proteins

The SARS-CoV-2 *Spike* protein, *3CLpro* (main protease, also called *Mpro*), and *nucleocapsid* protein are vital targets for therapeutic research because they are essential to the virus's capacity to infect, reproduce, and assemble inside host cells. The virus's glycoprotein Spike protein binds to the angiotensin-converting enzyme 2 (ACE2) receptor on the surface of host cells to mediate viral entrance. The two subunits that make up this complex are S1, which has the receptor-binding domain (RBD) necessary for binding to ACE2, and S2, which promotes membrane fusion and lets viral RNA into the host cell. This treatment or vaccine could, therefore, place an end to the infection even before it initiates by blocking the virus from attaching and invading human cells. Many COVID-19 vaccines and monoclonal antibody therapies that attempt to neutralize the virus are founded on this concept. Additionally, since mutations of the *Spike* protein correlate with changes in the transmissibility and immune evasion capabilities of the virus, targeting this protein might offer protection against other SARS-CoV-2 strains.

The SARS-Cov-2 protease enzyme *3CLpro* catalyzes the processing of polyproteins from the viral RNA for generating functional proteins needed to replicate and assemble the viruses. Inhibition of the activity of the *3CLpro* enzyme interrupts the replication cycle of the virus. New virions must also be generated for the production of viral proteins. Since *3CLpro* is a protein exclusive to the virus and does not have any counterpart in humans, drugs that target this protein are likely to be very specific and may have fewer side effects on the host. Additionally, because this family of viruses has very similar proteases, drugs targeting *3CLpro* may also offer broad-spectrum antiviral efficacy against other members of the coronaviruses.

Another important SARS-CoV-2 component is the *nucleocapsid* protein, a protein that interacts with the viral RNA genome in creating a ribonucleoprotein complex. This complex is essential both during the virus's packaging and during its replication. The drugs may limit the reproduction of viruses and proliferation if they cause the disruption of the ability of the virus to properly package its RNA or build new virions by targeting the *nucleocapsid* protein. Further, the *nucleocapsid* protein plays an important role in regulating the host's immune response and its inhibition may enhance the host's immunological power to counter the virus. Since it is less prone to mutation compared with *Spike* protein, this protein acts as a stable target for the synthesis of antiviral drugs. In addition, the life cycle of SARS-CoV-2 involves the *Spike, 3CLpro, and nucleocapsid* protein. These proteins also show promise as therapeutic targets for blocking the virus and halting the development of COVID-19.

### 3.1.2. Dataset

The Zinc database was used to extract both the drug candidates approved by the FDA and those approved globally [32]. Initially, a dataset including 5903 medications was acquired from https://zinc20.docking.org/ on October 02, 2023. With over 1.4 billion compounds, the Zinc database is openly accessible to the general population. Terabytes of data are downloaded from this website each week. The majority of compounds, more than 90%, are confirmed.

### 3.1.3. Data Preprocessing

The Zinc database, which consists of initially 5,903 approved drugs in SMILES format, was stepped through several conversion steps to allow for further analysis. The SMILES strings are converted into SDF format using the OpenBabel-2.4.1 software [33]. Then, the SDF files are converted into PDBQT format, which will enable the use of the drugs for calculating their binding affinity with the target protease. Files that could not be converted without errors were excluded from the dataset. After such a pre processing step, the dataset reduces to only 5,537 drugs.

### 3.1.4. Molecular Docking

On October 02, 2023, the crystal structures of the *Spike* (PDB ID: *7LM9*), *3CLpro* (PDB ID: *7JSU*), and *nucleocapsid* (PDB ID: *7DE1*) proteins were downloaded from the RCSB Protein Data Bank. Ligands from each of the structures were stripped to optimize the protein models. Water molecules and alternative side chains were removed from each structure; however, polar hydrogen atoms were retained. Macromolecules were added with Kollman charges. A GridBox of 30 x 30 x 30 and spacing of 1 Å were used to include the active sites of these proteins. The centers of the x, y, and z coordinates for *7LM9, 7JSU,* and *7DE1* are placed at 32.951, -13.678, -11.595; -11.046, 12.826, 67.749; and 29.275, 18.899, 17.009, respectively. Molecular docking was carried out using Auto-Dock VINA version 1.2.0 with default settings [26]. Ligands were prepared in PDBQT format using OpenBabel software. The binding affinities were calculated in kcal/mol. In all cases, the interaction with the lowest binding energy was the most favorable pose for the ligand binding. To illustrate our method, we showed in **Figure 4** crystal structures of *7LM9, 7JSU,* and *7DE1* determined at resolutions of 1.53 Å, 1.83 Å, and 2.00 Å, respectively.

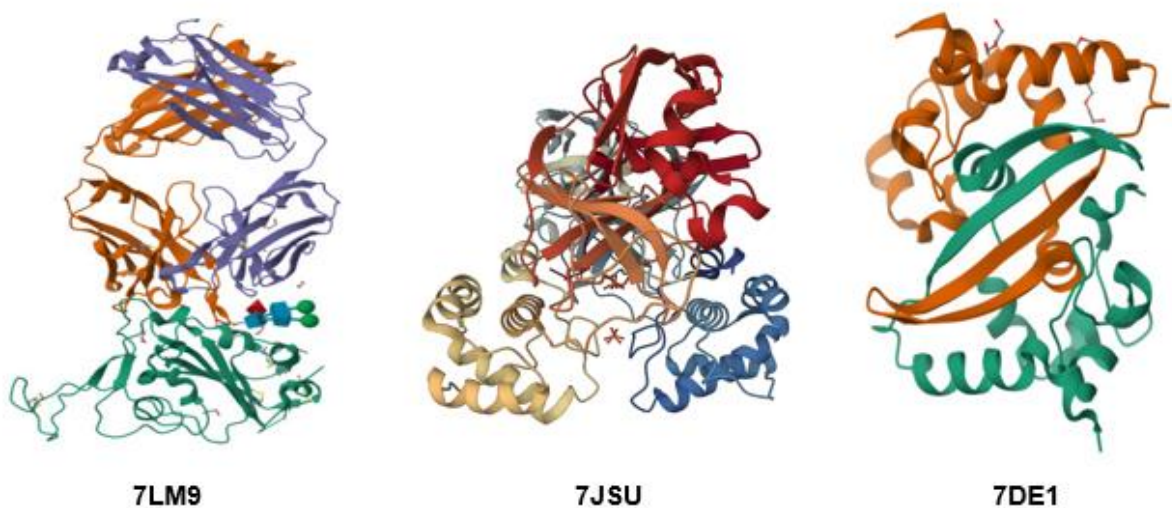

**Figure 4. Crystal structures of *Spike 7LM9, 3CLpro 7JSU and Nucleocapsid 7DE1*.**

## 3.2. Module B: QSAR Modeling

We have made a selection of numerous ML based regression models such as DTR, GBR, ETR, MLPR, and KNNR for QSAR modeling. These models predict QSAR between the physical properties of chemical substances and unknown biological activities. Utilizing these ML-based models, a relation has been determined between the biological activities and structural characteristics of identified chemical compounds. Physical qualities specify physicochemical properties, whereas biological activities specify pharmacokinetic properties. Application of molecular descriptors of complexes allows for an evaluation of changes in the structural features that impair biological function.

### 3.2.1. Data Cleaning

The data set is highly cleaned to remove all the duplicate data before the start-up of the machine learning models data preparation procedure. Additionally, drugs with missing binding affinity values are filtered from the data set to provide quality information. The final cleaned data set has a total of 4639 drugs and is ready for further analysis and modeling processes. Elimination of redundant data and exclusion of those drugs without binding affinity values ensures the accuracy and integrity of the dataset and its use.

### 3.2.2. Feature Extraction

A vector of fingerprint descriptors represented molecular components of drug compound. Before the calculation of descriptors, a built-in PaDEL- Descriptor tool [27] was applied for standardization of tautomer and desalting. Here in the current investigation, MACCS fingerprint descriptor was used to predict the binding affinities of drug molecules. The MACCS fingerprint encodes structural characteristics of the chemical compounds in a compressed and highly understandable form, so it is particularly useful for predicting binding affinity in the drug discovery process. It gives a binary, yes or no, whether certain substructures, such as functional groups, rings, or types of bond are present within a molecule or not. MACCS fingerprints allow for the efficient comparison of similarities between a query molecule and known ligands with established binding profiles in the prediction of binding affinity. We can find compounds that share structural properties with ligands known to bind efficiently to a particular target protein by comparing these fingerprints with similarity metrics.

The similar binding behavior of molecules with great structural similarity makes this similarity-based approach a useful proxy for estimating binding affinity. MACCS fingerprint, which allows the comparison to be carried out rapidly and, therefore particularly useful for virtual screening provides the ability to analyze vast chemical libraries to try and find potential candidates with favorable binding affinities. Additionally, by clustering compounds based on their MACCS fingerprints, those can be grouped that might interact with similar targets, streamlining the selection of candidates for more detailed experimental validation. While MACCS fingerprints provide an efficient way to identify potential binders.

### 3.2.3. Decision Tree Regression (DTR) Model

The aim of the work was to develop regression models which could predict the continuous response variable that is, binding affinity with high accuracy using several predictor factors including fingerprint descriptors. Several machine learning techniques are created for QSAR modeling in order to accomplish this goal. Because it performs better in predictions than the other models, the DTR approach is preferred. One kind of predictive model used in machine learning is called a decision tree-based predictive model, or DTR [34]. A decision tree, sometimes called a classification tree or regression tree, is also known as a graphical model, which itself could resemble a flow chart, having nodes or bubbles, branches, and leaves. The interior nodes of the decision tree each represent a test for a particular attribute; the branch shows the test's outcome, and each leaf node shows a prediction or class label. For a DTR, the value of a leaf node is the continuous value, such as the mean or the median of target values in training samples associated with the same leaf node. In DTR, feature space is divided into subsets recursively based on the optimization of the feature space to minimize the variance reduction of the target variable. This procedure keeps going till the halting requirement is satisfied. Its interpretability, resistance to over-fitting, and capacity to manage non-linear interactions between the features and the target variable are only a few of its benefits.

As mentioned in section 3.2.1, the input dataset contains 4639 drug compounds in total. This dataset has an 80:20 ratio between its internal and external datasets. Using 5-fold cross validation, the model's performance is robust and trained using the internal dataset. MACCS fingerprint molecular descriptor, which is described in section 3.2.2, is employed as the feature set for this purpose. The model's performance is evaluated using the external dataset.

In the following evaluation, the working of the resulting regression models is measured based on the two statistical measures: the coefficient of determination ($R^2$) and the root mean square error (RMSE). The value of $R^2$ measures the fraction of variance in the dependent variable which the independent variables can explain. A value of 1 implies a perfect fit, whereas a value of 0 can be considered a bad fit. However, RMSE offers a measurement of the prediction model's relative inaccuracy. A comparison study is performed in order to evaluate the performance of several regression models. We used MACCS fingerprint as the feature set for the comparative analysis.

## 3.3. Module C: Molecular Docking and Drug Analysis

In module C, two types of analysis are carried out. One is molecular docking analysis and the other are drug analysis.

### 3.3.1. Molecular Docking Analysis

In molecular docking analysis using AutoDock Vina, the process begins by providing the protein structure and ligand, which are used to predict the most favorable binding orientations and interactions within the protein's active site. The primary result of this analysis is the calculation of binding affinity, expressed as the free energy change ($\Delta G$) in kcal/mol. Lower binding energy values signify stronger and more stable ligand-protein interactions, with the pose displaying the lowest energy generally considered the optimal binding mode. In order to determine how the ligand interacts with the protein's active site, AutoDock Vina generates several binding poses for the ligand, which are then ordered according to their energy scores. Usually, the top-ranked poses are investigated.

In addition, the consistency of projected ligand poses in relation to a reference structure or experimentally acquired data is evaluated by computing the root mean square deviation (RMSD), where lower RMSD values denote more trustworthy docking predictions.

At this point, by showing how the ligand fits into the active site and interacts with the residues around it, visualization of the binding site contributes to a more profound understanding of the subject matter. This stage helps determine the orientation of the ligand and possible binding pockets. Such is the ability of the AutoDock Vina scoring tool that its docking findings can be ranked according to their binding affinities, and it is therefore possible to select the best position for further investigation. The entire given docking analysis will prove informative regarding the ligands-protein interaction, thus showing a clear path toward logical design according to presumed binding affinities and modalities of activators or inhibitors.

### 3.3.2. Drug Analysis

While contemplating drug repurposing by using in silico research to understand whether licensed drugs might be used for novel therapeutic uses, a thorough physicochemical evaluation is required. Such factors are crucial in defining the drug's permeability, solubility, bioavailability, and general pharmacokinetic behavior. Among the common physicochemical factors analyzed is molecular weight, which influences the general absorption profile of the drug and its ability to cross biological membranes. The affinity of the drug for the lipid versus the aqueous environment, or lipophilicity is also indicated as log P, and impacts its solubility and distribution. The count of HBD and HBA determine the aqueous solubility of the drug and its ability to interact with the biological target.

Some examples of such additional features that explain the structural complexity of the drug and its ability to be involved in hydrophobic interactions include the existence of rings and the presence of heavy atoms. Knowledge of heteroatoms, which are atoms other than carbon and hydrogen, gives an understanding of the drug's ability to engage in hydrogen bonding and other electrostatic interactions. The proportion of sp3 hybridized carbons (SP3) is ultimately analyzed to estimate the three-dimensionality of drugs and their possible impact on binding affinity and metabolic stability. Finally, it can predict pharmacokinetic behavior, safety, and efficacy of repurposed pharmaceuticals, which makes evaluation easier for new therapeutic applications.

## 4- Conclusion

Our study was successfully able to identify drugs already available which could be repurposed as the drugs for the treatment of the disease COVID-19 by checking the possible target proteins, which are *Spike, 3CLpro,* and *Nucleocapsid*. Using a computational approach that can integrate molecular docking and machine learning, we were able to identify five drugs that have high binding energies ranging between -19.7 and -12.6 kcal/mol, besides being favorable in terms of binding energies, making them promising drugs for the treatment of COVID-19. Our proposed DTR model produced better predictions of binding affinities than others in comparison with the other regression methods with R² values of 0.95, 0.97, and 0.93 and RMSE values of 1.66, 1.57, and 1.49 for the *Spike, 3CLpro,* and *Nucleocapsid* proteins, respectively, using MACCS fingerprints.

We also determined the physicochemical properties of selected compounds to establish their action mechanisms in the biological environment, hence determining the drugs' efficacy and safety profile. Our results showed these compounds to be adequately potent inhibitors of multi-target proteins associated with COVID-19. This study hence provides a basis for further trials in order to prove the validity of such compounds as potential COVID-19 therapeutics. Follow-up research, including in vitro and in vivo studies, is necessary to confirm their effectiveness. Moreover, exploring the possible synergistic effects of these compounds in combination with other treatments could yield valuable insights. Our methodology could also be applied to other viral diseases and conditions beyond virology, making drug repurposing a promising strategy. Ultimately, this study underscores the significance of utilizing computational tools to expedite the drug discovery process, particularly during emerging pandemics and urgent health challenges.

**Conflicts of Interest:** The authors declare no conflict of interest.